\documentclass[preprint2]{aastex63}

\usepackage{graphicx}
\usepackage[suffix=]{epstopdf}
\usepackage{natbib}
\usepackage{amsmath}
\usepackage{url}
\usepackage{xspace}

\providecommand{\e}[1]{\ensuremath{\times 10^{#1}}}

\begin{document}

\title{The Rise and Fall of the Eclipsing Binary HS Hydrae}

\correspondingauthor{James. R. A. Davenport}
\email{jrad@uw.edu}

\author[0000-0002-0637-835X]{James. R. A. Davenport}
\affiliation{Astronomy Department, University of Washington, Box 951580, Seattle, WA 98195, USA}

\author{Diana Windemuth}
\affiliation{Astronomy Department, University of Washington, Box 951580, Seattle, WA 98195, USA}

\author{Karen Warmbein}
\affiliation{Astronomy Department, University of Washington, Box 951580, Seattle, WA 98195, USA} 

\author[0000-0002-0716-947X]{Erin L. Howard}
\affiliation{Department of Physics \& Astronomy, Western Washington University, Bellingham, WA 98225}

\author[0000-0002-2762-4046]{Courtney Klein}
\affiliation{Department of Physics and Astronomy, University of California Irvine, CA 92697, USA }

\author[0000-0002-7961-6881]{Jessica Birky}
\affiliation{Astronomy Department, University of Washington, Box 951580, Seattle, WA 98195, USA}

\begin{abstract}
HS Hydrae is a short period eclipsing binary ($P_{orb}=1.57$ day) that belongs to a rare group of systems observed to have rapidly changing inclinations. This evolution is due to a third star on an intermediate orbit, and results in significant differences in eclipse depths and timings year-to-year. \cite{zasche2012} revealed that HS Hydrae's eclipses were rapidly fading from view, predicting they would cease around 2022. Using 25 days of photometric data from  Sector 009 of the Transiting Exoplanet Survey Satellite (TESS), we find that the primary eclipses for HS Hydrae were only $0.00173\pm0.00007$ mag in depth in March 2019. This data from TESS likely represents the last eclipses detected from HS Hydrae. We also searched the Digitization of the Harvard Astronomical Plate Collection (DASCH) archive for historic data from the system. With a total baseline of over 125 years, this unique combination of data sets – from photographic plates to precision space-based photometry – allows us to trace the emergence and decay of eclipses from HS Hydrae, and further constrain its evolution. Recent TESS observations from Sector 035 confirm that eclipses have ceased for HS Hya, and we estimate they will begin again in 2195. 
\end{abstract}

\section{Introduction}

Eclipsing binary stars (EBs) are the most important benchmark systems for calibrating models of the fundamental properties of stars \citep{stassun2009}. By observing both the eclipses with photometry and radial velocity variations throughout the orbit with spectroscopy, we can directly estimate the masses, radii, and surface brightness for both stars. In the modern era of ultra-precise, space-based photometry, EBs enable incredibly precise and accurate measurements of the fundamental properties of stars \citep{miller2020}. Such precision also allows us to search for low-amplitude signals from the EB, which can be due to companion stars or second-order physical effects \citep{bloemen2013}.

A rare class of EB systems are those with observable changes in their inclinations or alignments over time. Such systems are often discovered as EBs with observable eclipse depths and/or eclipse times that vary over many years. In the case of changing inclination systems, this results in a window of time (typically decades) where eclipses are visible to our line of sight, and longer time spans (often centuries) where they are not. Some binaries with historical eclipse data are now classified as {\it former} EBs due to this precession \citep[e.g. QX Cas;][]{bonaro2009}.

This evolution is typically ascribed to the influence of a gravitationally bound third star. These rapidly evolving systems are therefore unique and important windows into the structure and dynamics of triple star systems \citep{soderhjelm1975}.
Wide-area sky surveys are now regularly producing large catalogs EBs \citep{jayasinghe2018}, and have started to produce catalogs of these dynamic systems, such as from the Magellenic Clouds via changing eclipse depths \citep{jurysek2018}
or precise eclipse timing variations 
\citep{borkovits2016}. However, 
only a few inclination-changing EBs have been well studied within the Milky Way \citep{guinan2012,jurysek2018}.

HS Hydrae has long been known to be an EB \citep{strohmeier1965}, and has been studied in detail as a normal binary system \citep{gyldenkerne1975}. \cite{torres1997} first identified the presence of a third stellar body in the system, estimating the  outer orbital period to be 190 days. \cite{zasche2012} found that HS Hya belongs to the rare class of precessing, inclination-changing EBs. \citet{zasche2012} further predicted eclipses would cease for HS Hya around 2022, and presented the need for additional precision follow-up for the system.

Here we present a comprehensive archival study of the inclination-changing EB, HS Hya.
We combine historical archival photometry from photographic plate archives, ground-based surveys, and new precision space-based light curves in \S\ref{sec:data}. 
In total we have assembled a unique data set that spans over 125 years. This data set maps the entire observable history of eclipses from HS Hya, which we model in \S\ref{sec:risefall}.
We believe that the TESS mission \citep{tess} serendipitously captured the final observable eclipses in 2019, and in \S\ref{sec:final} we predict that HS Hya will resume eclipsing around 2195.
In \S\ref{sec:others} we review similar systems that TESS is uniquely suited for studying, and finally provide a brief summary and discussion in \S\ref{sec:discuss}.

\vspace{0.5in} 
\section{A Century of Data}
\label{sec:data}

HS Hya is a bright system at $V\sim$8.1 mag, and relatively nearby with a distance of $103\pm7$ pc \citep{gaia_edr3}. As such, considerable data exists for HS Hya in photometric archives and historical literature. Bright stars are also ideal targets the current generation of exoplanet-hunting missions with high precision and high cadence photometry. In this section we detail the wide range of data gathered for for our study of HS Hya, reaching back over a 125 year baseline.

\subsection{Archival Data}
\label{ssec:archive}

\begin{figure}[!t]
\centering
\includegraphics[width=3.5in]{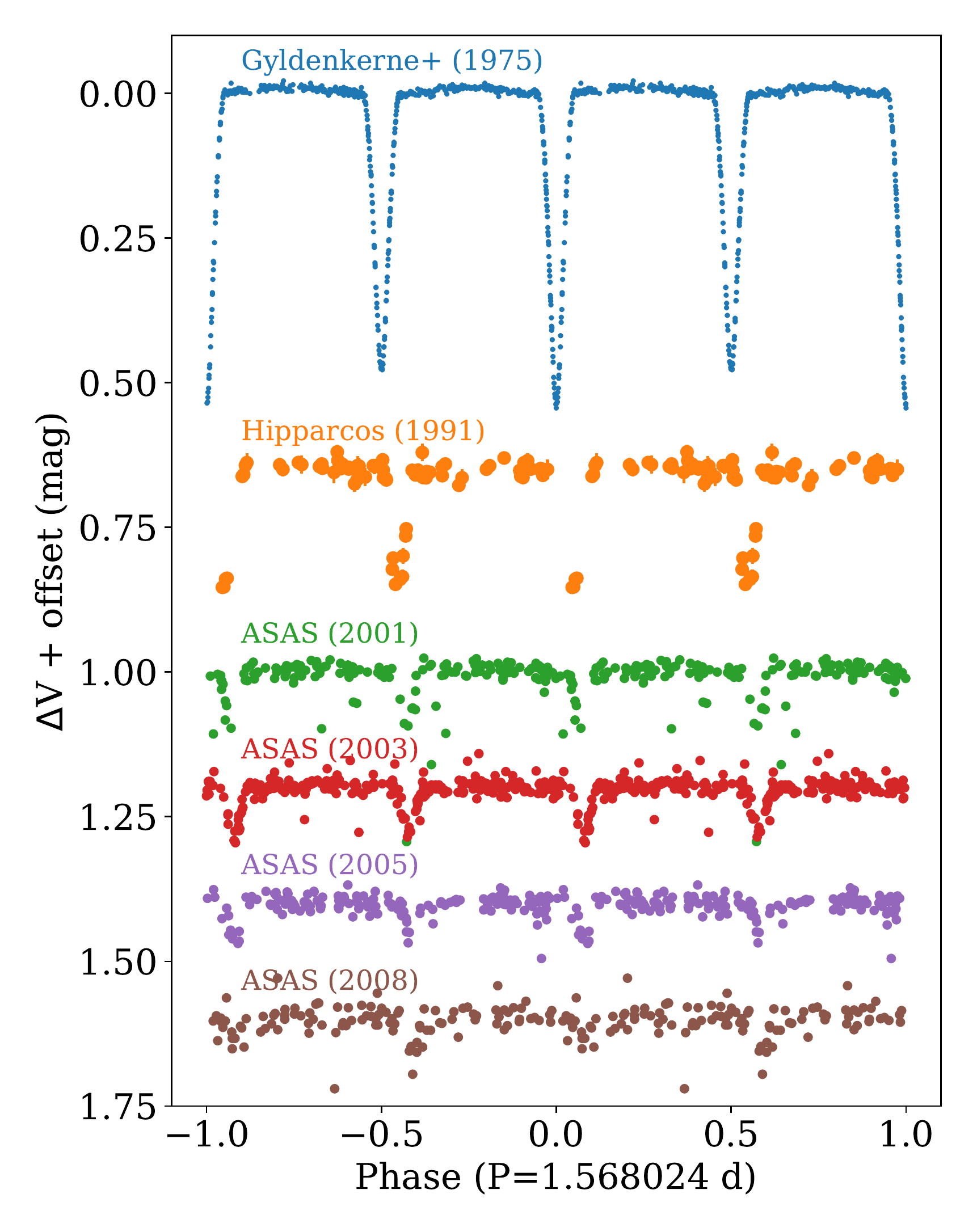}
\caption{
Phase-folded light curves for HS Hydrae from various photometric studies and archives. As in \citet{zasche2012}, a clear decay in  eclipse depth and precession of eclipse phase can be seen from 1975 to the faintly detectable eclipses in 2008.
}
\label{fig:archive}
\end{figure}

As \citet{zasche2012} note, several sources of archival data for HS Hya exist.
In this section we briefly review the history of observations for the system, starting by reviewing the archival data used by \citet{zasche2012}, shown in Figure \ref{fig:archive}.

HS Hya was first identified as an eclipsing binary in 1964 as part of a survey of bright variable stars \citep{strohmeier1965}. Though the orbital period was not correctly identified, the approximate depth of the primary eclipse can be estimated from their Figure 5. We use the approximate eclipse depth in later analysis, but are unable to phase-fold their data into our ensemble in Figure \ref{fig:archive}.

\citet{popper1971} published the first radial velocity curves, as well as the correct orbital period (1.568024 days), and raised the need for more photometric data for HS Hya. \cite{gyldenkerne1975} then obtained 533 observations over 12 nights of extremely precise Str\"{o}mgren $ubvy$ photometry for HS Hya, shown in Figure \ref{fig:archive} (blue points). We use the primary eclipse ephemeris from \citet{gyldenkerne1975} and orbital period from \citet{popper1971} throughout our work:

\begin{equation}
    \phi_{pri} = 2441374.5934 + 1.^{\kern-0.25em {\rm d}}568024\times E
\end{equation}

\noindent 
where $\phi_{pri}$ is the orbital phase of the primary eclipse, the period is in days, and the ephemeris is in Julian date. 
The Hipparcos mission \citep{perryman1997} was able to capture 79 additional epochs of high quality photometry for HS Hya (Figure \ref{fig:archive}, orange points).

As \citet{zasche2012} show, the All Sky Automated Survey \citep[ASAS;][]{asas, pojmanski2002} provided many years of photometric monitoring for HS Hya. We obtained the latest version of the ASAS $V$-band light curve for HS Hya, which had a total of 662 epochs of usable photometry with the {\tt GRADE=`A'} quality flag set. We grouped the ASAS data into four portions of roughly equal time resolution. The mean year within each time range is noted in Figure \ref{fig:archive}.

We also searched for other modern catalogs for useful data to compliment the ensemble listed above. We queried both the individual epoch (L1b) data from the infrared WISE mission \citep{wise}, as well as the optical ground-based ASAS-SN survey \citet{kochanek2017}, which has been used to generate large samples of eclipsing and variable sources \citep[e.g.][]{jayasinghe2020}. Light curves for HS Hya were available from both catalogs, but the photometric precision was not high enough in either to detect the eclipses, where were already below 0.1 mag in amplitude in 2008.

\subsection{DASCH}
\label{ssec:dasch}

We were then able to substantially extend the baseline of the archival data used in our analysis. The Digital Access to a Sky Century at Harvard (DASCH) is an effort to create a uniform digital catalog of photometric measurements from astro-photographic plates taken between 1885 to 1992 \citep{laycock2010,tang2013}. Each plate is digitized by back-illuminating it and scanning using a CCD camera. That image is then calibrated based on the plate's sensitivity, and PSF photometry is computed for each source in the image.

\begin{figure}[!t]
\centering
\includegraphics[width=3.5in]{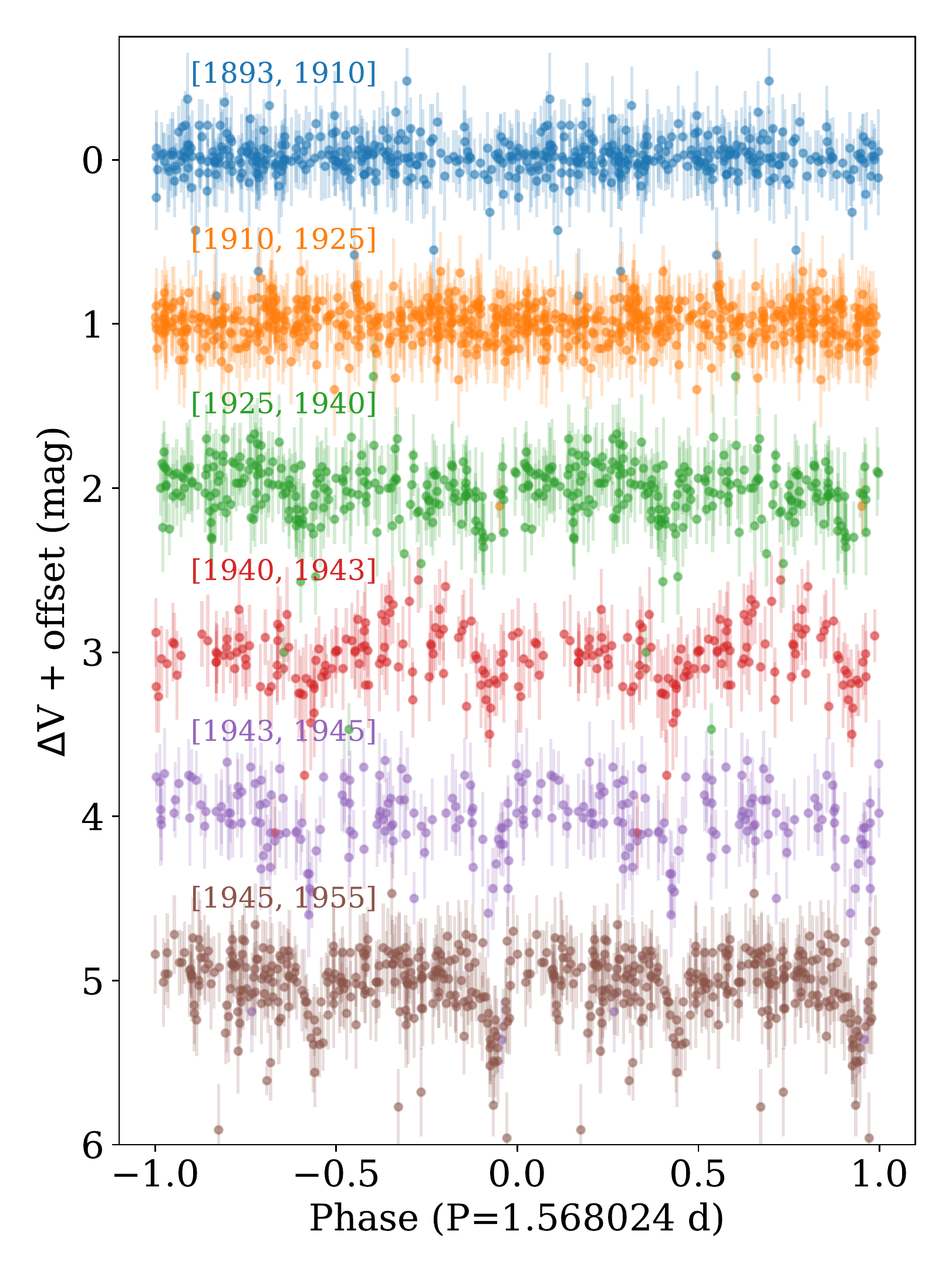}
\caption{
Phase-folded historical light curves from DASCH for the inclination-changing, eclipsing binary system, HS Hydrae in six bins of time. Time ranges for each bin are noted in brackets. While we believe eclipses started in $\sim$1897, {\it measurable} eclipses from photometric plates were not apparent for the system until the 1920's.
}
\label{fig:dasch}
\end{figure}

Setting an upper limit for the photometric uncertainty of $\sigma<0.3$ mag, we found 1390 usable observations for HS Hya from the DASCH archive. This light curve spanned from 1893 to 1989, and the phase-folded data is shown in Figure \ref{fig:dasch}. These data had a mean photometric uncertainty of 0.2 mag, which was remarkably stable throughout the time series. As with the ASAS data, we binned the light curve into ranges of time with roughly equal numbers of observations to illustrate in Figure \ref{fig:dasch} the changing system inclination (or eclipse depth) over time. We discard data after 1955, as the available archive of plates became very sparse in time. In total we only omitted 121 ``good'' epochs between 1955 and 1989.

No sign of the eclipsing nature for HS Hya is apparent for the first several decades of DASCH data in Figure \ref{fig:dasch}. The EB begins to reveal itself clearly in the early 1920's, as the eclipse depth begins to reach above 0.1 mag. By the final time window in our DASCH data, between 1945 and 1955, the primary eclipse has a very large depth of at least 0.5 mag.

\subsection{TESS}

\begin{figure*}[!t]
\centering
\includegraphics[width=7in]{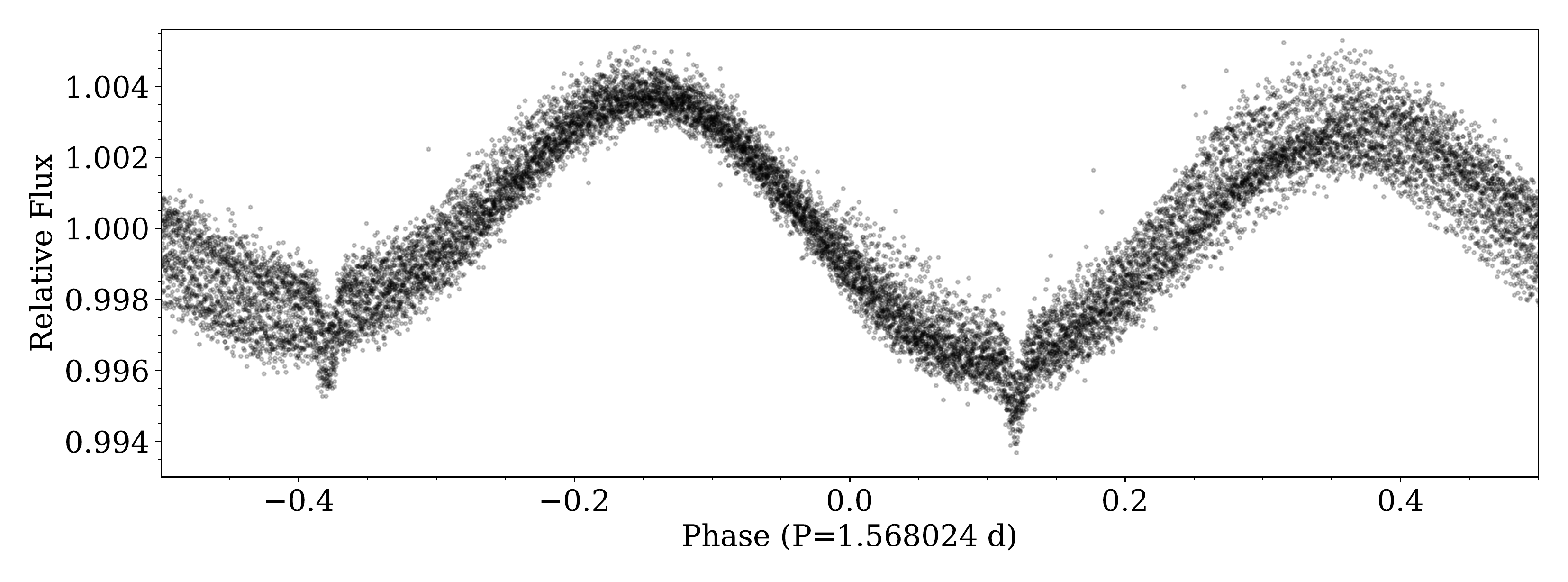}
\caption{
TESS Sector 9 phase-folded light curve for HS Hydrae. As in Figures \ref{fig:archive} and \ref{fig:dasch}, the TESS data have been phased using the orbital period and ephemeris from \citet{gyldenkerne1975}. Very small amplitude primary (Phase$\sim$0.1) and secondary (Phase$\sim$0.6) eclipses are visible, as well as out-of-eclipse ellipsoidal variation, and possible variability due to starspots. As this system is rapidly precessing, we believe these may be the final observable eclipses for HS Hya for almost two centuries.
}
\label{fig:tesslc}
\end{figure*}

The Transiting Exoplanet Survey Satellite \citep[TESS;][]{tess} is a NASA mission currently studying bright stars from over 80\% of the sky. In its primary mission, TESS provides nearly continuous light curves with between $\sim$27 day and 1-year baselines. TESS captures 30-minute cadence imaging for its entire field of view, and generating 2-minute cadence light curves for $\sim$20,000 pre-selected targets simultaneously.

HS Hya (also known as TIC 434479378) was observed in Sector 009 of the TESS primary mission, and was selected to receive 2-minute (short) cadence data shown in Figure \ref{fig:tesslc}. The observations for Sector 009 spanned 2019-02-28 through 2019-03-26, with the characteristic TESS mid-Sector gap for data down-link. The  apparent magnitude of HS Hya is ideal for generating an ultra high precision light curve, as shown in Figure \ref{fig:tesslc} with a mean photometric error in relative flux of 3\e{-4}.

The phase-folded TESS light curve for HS Hya is dominated by a $\sim$1\% flux modulation that is remarkably stable throughout the Sector. This periodic modulation is due to the ellipsoidal variation of the system, as the binary is on very short period orbit.
Small variations in the ellipsoidal variation are likely due to small amplitude systematic errors in detrending the TESS data across the entire quarter.

Very small amplitude eclipses are visible in the phase-folded TESS light curve in Figure \ref{fig:tesslc}. To compare the eclipse depths with the previous measurements, we convert the relative fluxes to magnitudes using the typical approach of $\Delta mag = -2.5 \log_{10} f_{rel}$. Using a simple Gaussian fit to the eclipse, we estimate the grazing eclipse of HS Hya had an amplitude of $0.00173\pm 0.00007$ mag in 2019.

As with many EBs observed by missions like Kepler and TESS, ultra high precision constraints are possible on both the fluxes due to the photometric precision (e.g. small amplitude features as shown here), and for the temporal information due to the dense light curve sampling (e.g. the phase of the eclipse). For HS Hya, the TESS data show the ellipsoidal variation does not align with the eclipses exactly in orbital phase. The primary eclipse in Figure \ref{fig:tesslc} leads the bulge by a very small but significant amount of the orbital phase. We find the eclipse leads the light curve minimum by $0.0114 \pm 0.0004$ in phase. 
The precision of this phase lag may be useful for constraining the tidal synchronization for such systems \citep{zahn1977, barnes2011}.

\section{Modeling 125 Years of HS Hya}
\label{sec:risefall}

Our photometric data for HS Hya spans a +125 year baseline, from 1893 to 2019. As Figures 1--3 show, there is significant evolution in both the amplitude and phase of the eclipses. The phase of the primary eclipse visibly moves from -0.2 in the early DASCH data (Figure \ref{fig:dasch}), to 0.11 in the TESS data (Figure \ref{fig:tesslc}). These effects are due to orbital precession and inclination evolution of the system owing to a third body in orbit around the eclipsing components, as described by \citet{zasche2012}. 

\begin{figure}[!t]
\centering
\includegraphics[width=3.25in]{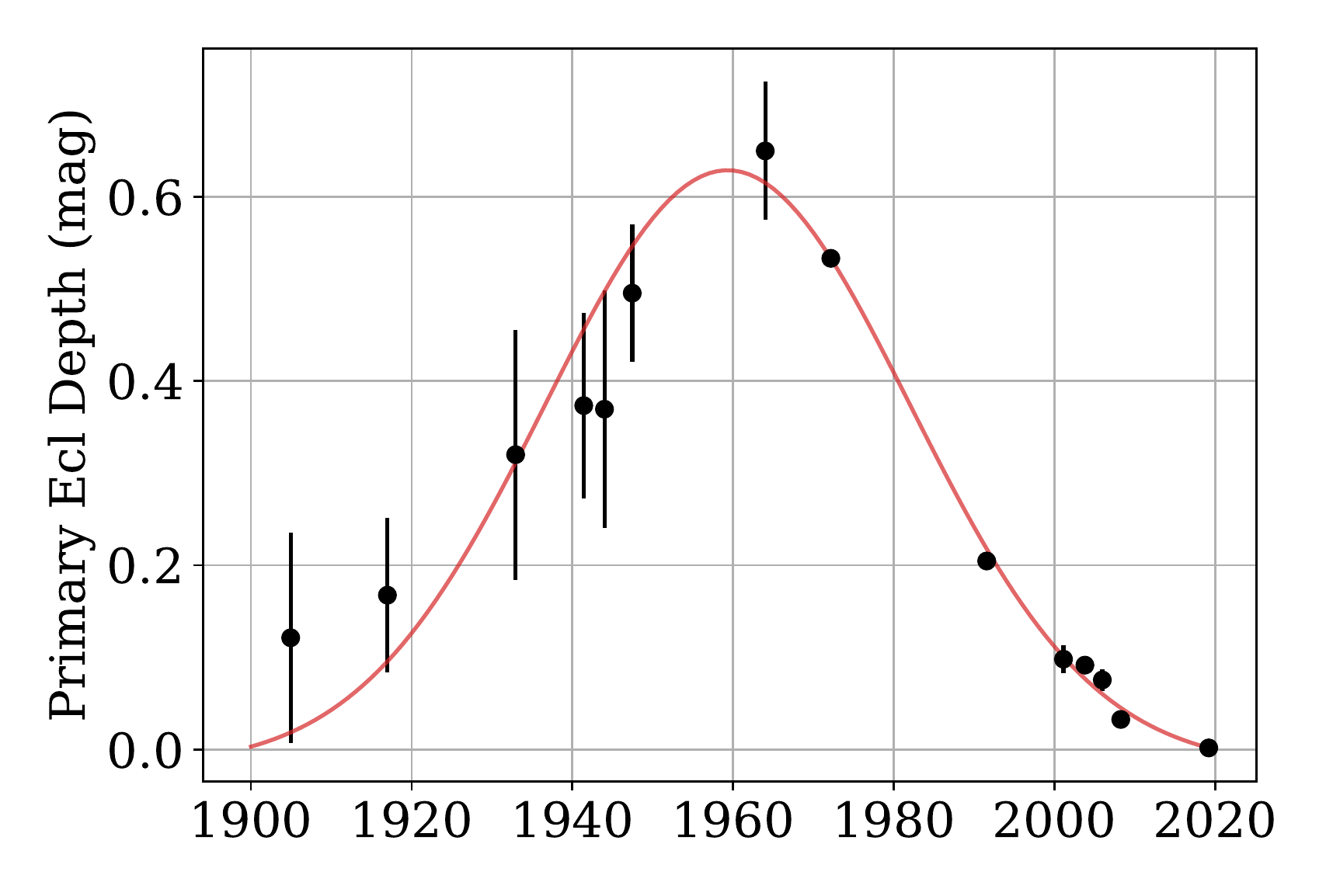}
\caption{
Eclipse depths as a function of time for HS Hydrae, combining data spanning over 125 years. Primary eclipses within each time span were fit using a simple Gaussian profile. A Gaussian profile (red line) was fit to illustrate the system's entire history as an eclipsing binary. We find the maximum eclipse depth occurred in $\sim$1959.
}
\label{fig:depth}
\end{figure}

To illustrate the ``rise and fall'' of eclipses from HS Hya, in Figure \ref{fig:depth} we show a phenomenological model of the observed eclipse depths over time. We fit the primary eclipse from the phase curves shown in Figures 1--3 using a Gaussian curve to estimate the eclipse depth within each window of time. Additionally we include the EB discovery epoch from \citet{strohmeier1965}, where the eclipse depth and uncertainty are estimated by eye, following \citet{zasche2012}. Figure \ref{fig:depth} shows the eclipses were barely visible near the beginning of the 20th century. By the 1940's, the EB is well detected. As the inclination evolves approximately linearly with time for these systems \citep{torres2001}, the resulting change in eclipse depth over time is, to first order, a Gaussian profile \citep[e.g. Fig 4 of ][]{drechsel1994}. This simple summary model of the eclipse depths over 125 years in Figure \ref{fig:depth} shows HS Hya was at maximum eclipse depth around $\sim$1959.

To determine the inclination of HS Hya over time, we then generated eclipsing binary models for each of the phase curves in Figures 1--3. We utilized the detached EB modeling framework {\tt KEBLAT}, developed by \citet{windemuth2019}. {\tt KEBLAT} leverages stellar evolution models \citep{bressan2012} to infer robust physical parameters for EBs using only photometry and without radial velocity data -- e.g. with data from missions like Kepler and TESS. We refer the reader to \citet{windemuth2019} for details about the photodynamical modeling, and describe below how we adapt {\tt KEBLAT} to fit light curves from disparate photometric surveys and model the time varying physical parameters. 

We use an iterative approach to fit the entire data set, which spans 12 epochs of V-band data and 1 epoch of TESS data (i.e. Figures \ref{fig:archive}--\ref{fig:tesslc}). We individually fit each epoch in a self-consistent way, and then cumulatively combine the epochs into one model to measure inclination and time of primary eclipse as a function of the $i^{th}$ mid-epoch, $E_i$. 
We hold masses and temperatures fixed to radial velocity-derived masses and temperatures from \citet{torres1997}, as they are more accurate, and we lack multi-band photometry required for {\tt KEBLAT} to infer masses via stellar isochrones alone. Note that we also do not fit for a ``third light'' component for the system.

We first initialized our {\tt KEBLAT} model using only the data from \citet{gyldenkerne1975}. We started with this epoch as it had the highest signal-to-noise eclipse data, and was close to the epoch of maximum eclipse depth for the system according to Figure \ref{fig:depth}. We set initial parameter values for the orbital period and time of primary eclipse ($P$ and $t_{\mathrm{pe}}$) to literature values; radii and flux ratio values to PARSEC stellar isochrone predictions based on the RV-derived masses \citep{bressan2012}; eccentricity vectors $e \sin \omega$, $e \cos \omega$) using primary and secondary eclipse widths and separations; and finally random small initial values to impact parameter $b$ and limb darkening coefficients. We then used {\tt lmfit}, a non-linear least squares optimizer, to solve for the free parameters corresponding to the \citet{gyldenkerne1975} epoch. 

We then successively added the remaining 11 epochs of V-magnitude data to our {\tt KEBLAT} model. We assume that binary period and eccentricity, as well as physical parameters associated with the binary components themselves (i.e. stellar radii, V-band flux ratio, limb darkening coefficients) do not evolve on the century timescale of our data set, and thus hold them fixed at the optimum values from the \citet{gyldenkerne1975} epoch. We allow each epoch's impact parameter (inclination) and time of primary eclipse to float in our combined V-band data set, and solve for the best-fit values via {\tt lmfit}. 

Finally, we include the TESS epoch in our cumulative model, and fit for impact parameter and time of primary eclipse for all epochs of data. We use the PARSEC isochrone model grid to solve for a self-consistent TESS-band flux ratio, based on the published masses and effective temperatures for the binary components. We note that uncertainties in this transformation may result in over-estimating the inclination precision for the TESS epoch. However we believe this impact is minimal, as the EB has a nearly equal-mass ratio, i.e. ratio of eclipse depths will not be strongly wavelength dependent.

We emphasize that {\tt KEBLAT} did not parameterize any time dependent evolution for HS Hya. Instead, we created a final single solution for our data, solving for the point-in-time inclination and time of primary eclipse for each epochal data set as free parameters. 
Since our goal was to examine the evolution of system inclination, we do not consider the system properties (i.e. radii, flux ratios, eccentricity vectors) that {\tt KEBLAT} estimated from the \citet{gyldenkerne1975} epoch are robust determinations of the true values. As such, we did not produce uncertainties on these parameters. We do note that these values are all close to more detailed modeling of the system from \citet{torres1997}, and for completeness we include them in Table \ref{tbl1}.

\section{The Final Eclipses}
\label{sec:final}

\begin{figure}[!t]
\centering
\includegraphics[width=3.25in]{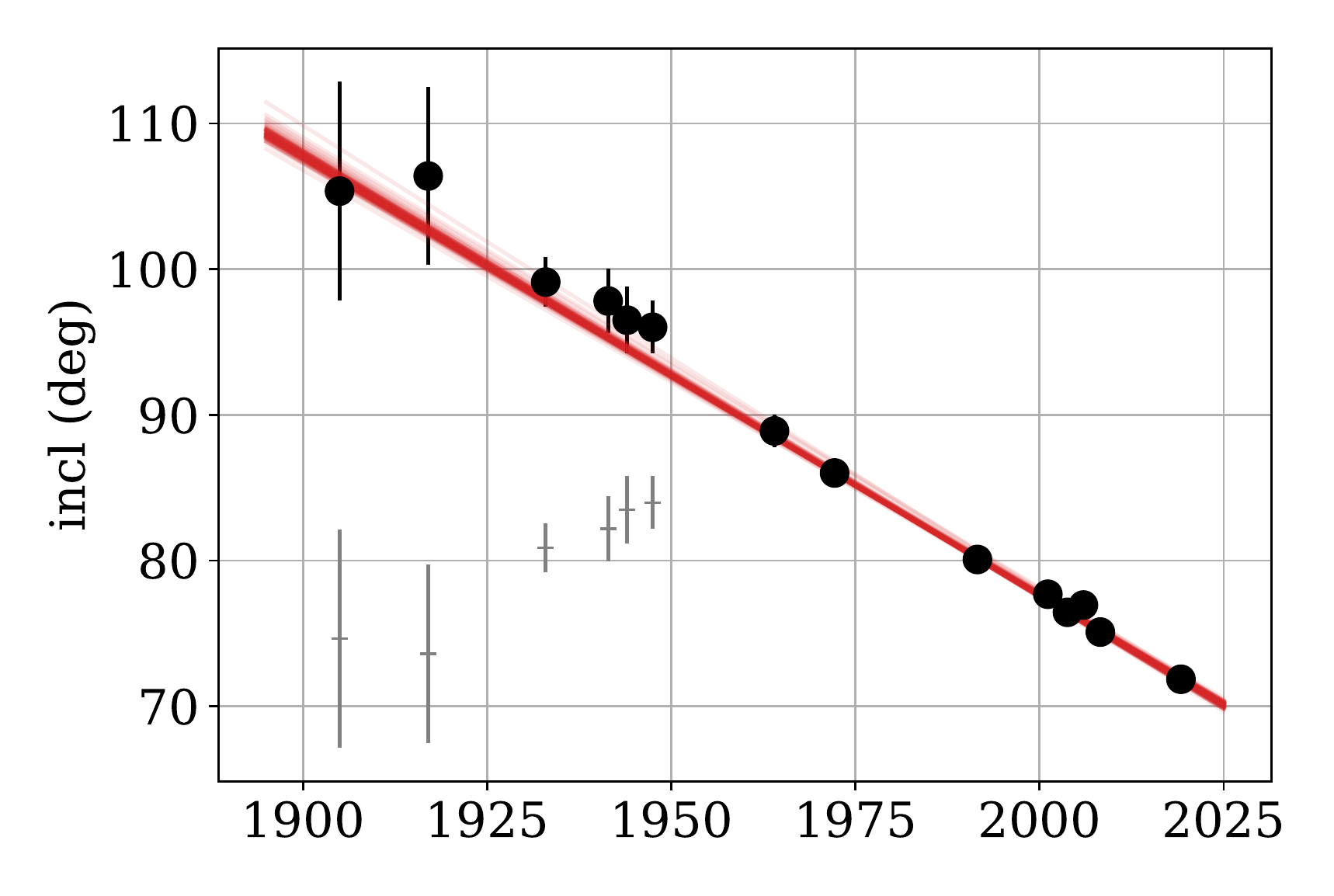}
\caption{
Derived inclination over time for HS Hya. For data before $\sim$1960, actual fits are shown (thin crosses), as well as inclinations flipped above $90^\circ$ to enable a linear fit throughout the observations.
Draws from the posterior distribution of this linear model are shown (red lines). The median model predicts that eclipses from HS Hya will be undetectable by TESS in September 2020, and that eclipses will cease in January 2021 when HS Hya reaches the critical inclination of $i=71.3^\circ$.
}
\label{fig:incl}
\end{figure}

From our {\tt KEBLAT} model we have derived estimates of the inclination HS Hya in 13 individual epochs spanning over a century. These inclinations, and their standard 1-$\sigma$ uncertainties, are shown in Figure \ref{fig:incl}. 

We find that the inclination for HS Hya changes nearly linear over the 125-year baseline, as expected from the evolution of other similarly evolving triple star systems \citep{torres2001}. The inclination angle for an eclipsing binary is typically defined between $0^\circ$ (pole-on, no eclipse) and $90^\circ$ (equator-on, deepest eclipse), and our {\tt KEBLAT} fits from \S\ref{sec:risefall} are defined within this range. However, in Figure \ref{fig:incl} we choose to flip the inclinations for the DASCH epochs {\it above} $90^\circ$, making this linear evolution more clear. Conceptually this corresponds to the system starting to eclipse as the secondary star begins to graze the ``top'' of the primary in the early 1900's (i.e. $i\sim108.7^\circ$), and eclipses end as the secondary star  grazes the ``bottom'' of the primary (i.e. the critical inclination of $i\sim71.3^\circ$).

We fit the inclination evolution in Figure \ref{fig:incl} using a linear model \citep{torres2001}. We initialize this fit using a least-squares regression to the inclinations and their uncertainties. This was then used to initialize an affine invariant Markov Chain Monte Carlo (MCMC) parameter exploration using {\tt emcee}. In Figure \ref{fig:incl} we show 50 draws from the posterior distribution of this MCMC fit, and we adopt the median values from the posteriors as our preferred inclination evolution model going forward. 

According to our {\tt KEBLAT} modeling, HS Hya will cease to have eclipses from our line of sight at the critical inclination of $i\sim71.3^\circ$. Our MCMC fit to the  evolution predicts that HS Hya will reach this inclination at year $2021.1\pm0.5$ (Feb 2021). Despite sampling a +125 year baseline, a half year uncertainty remains in the precise time. HS Hya is scheduled to be observed again during Cycle 3 of the TESS mission, in Sector 035 (Feb--Mar 2021). While it is possible the system will still be eclipsing at this time, the eclipse amplitudes will be so small that it is unlikely even TESS will be able to detect them as distinct from the ellipsoidal modulations. We therefore believe the TESS Sector 009 data in Figure \ref{fig:tesslc} are the final observable eclipses for HS Hya during this era.

Our linear inclination evolution modal predicts that HS Hya will reach the critical inclination again in the year $2195\pm3$. The ellipsoidal variations will continue to be observable by facilities like TESS for many decades to come. Their amplitudes will  decrease as the system continues to precess, which can be used to continue tracing the orbital evolution of the system using tools designed for modeling tidally distorted EBs such as {\tt PHOEBE} \citep{prsa2016}.

Rather than a linear model, \citet{zasche2012} estimated the ``nodal period'' of the system using a sinusoidal function fit to $\cos(i)$, following the approach \citet{drechsel1994} used for IU Aur. They found $P_{nodal}\sim631$ yr for HS Hya. Since only a small portion of the nodal period for HS Hya is observed in the past 125 years, sinusoidal functions are under-constrained by the data, as \citet{zasche2012} note. For completeness we have replicated this approach using our longer observation baseline for HS Hya. A sinusoidal least-squares fit to the $\cos(i)$ estimates from {\tt KEBLAT} yields $P_{nodal}=973$ years. Our linear inclination evolution model can also be restated as a ``tumble'' period for the system (i.e. how long it takes for the inclination to change by $360^\circ$). We find this tumble period to be $P_{nodal}=1194\pm20$ years for HS Hya.

\section{The Third Star}
\label{sec:3rd}
Throughout this work we have focused on the observed properties of the eclipsing component of HS Hya, while largely neglecting the third stellar body in the system. \citet{torres1997} have produced the only substantial characterization of star $C$, estimating it to have an orbital period of $P=190.529\pm0.061$ around the eclipsing $AB$ stars, and a spectral type of M0 (0.5$M_\odot$). The orbit for star $C$ was also estimated to be nearly co-planar to the inner stars' orbits, though no eclipses from star $C$ were reported.

Using our 125-year archive of photometry, we searched for any signs of eclipses from the third star in HS Hya. We searched each of the ground-based archival light curves shown in Figures \ref{fig:archive} and \ref{fig:dasch} for any periodic signals around $P=190$ days, using both Lomb-Scargle \citep[e.g.][]{gatspy} and Box Least Squares \citep{kovacs2002} periodogram techniques. 
No eclipses besides the expected $P=1.5$d were found in any of the ground-based datasets from \S\ref{ssec:archive}, though we note the DASCH data did show some low-amplitude signal in the Lomb Scargle periodogram broadly centered around 190 days. The origin of this signal was unclear, however, as it could occur due to e.g. half of the natural 1-year alias in the data.

The TESS light curve for HS Hya (Figure \ref{fig:tesslc}) only covers a small portion of the 190d orbital period for star $C$, and no extra eclipse features were detected for the system. In total the TESS data covered 13.5\% of the outer orbital period. Using the ephemeris for the predicted time of periastron for the third body by \citet{torres1997} (HJD=$2448063.8\pm3.6$ d), TESS narrowly misses the predicted periastron time for the third body, with the light curve starting 0.32 days after. There remains substantial uncertainty in the time of periastron without an eclipse. As HS Hya continues to dynamically evolve, and the third star is predicted to be nearly co-planar with the primary (formerly eclipsing) components, an eclipse of the inner stars by the third body is a possibility. Unfortunately we cannot predict from our data when such an event may occur, nor if it has occurred over the past century given the sparse sampling of our historic light curves.

Besides photometric (e.g. transits) and spectroscopic detection, the third star in HS Hya could potentially be characterized from precision astrometric monitoring. As \citet{stassun2021} note, for unresolved binaries the Gaia ``RUWE'' statistic may be a useful metric for inferring motion of the system's photocenter, which could occur due to the presence of a third star. RUWE characterizes the error of the astrometric solution for a system over the available data. HS Hya has a value from Gaia EDR3 of RUWE=2.38, far greater than the nominal range of 1--1.4 expected for single stars. This is an excellent indication that the eclipsing $AB$ components are moving due to the tertiary star. \citet{stassun2021} further produced an empirical relation between RUWE and the expected photocenter motion from known eclipsing systems. Using this relation, we find HS Hya has an estimated photocenter motion from the Gaia EDR3 RUWE of 0.58 mas. This is larger than the expected photocenter shift of the system from the small third star alone of $\sim$0.2 mas, but we note that the RUWE value for HS Hya is larger than any in the empirical relation provided by \citet{stassun2021}. 

In Table \ref{tbl1} we summarize the relevant system properties for HS Hya from the literature and this work, including values for the third star.

\begin{deluxetable}{lcc}[]
\tablecolumns{3}
\tablecaption{
    System Properties
    \label{tbl1}
}
\tablehead{
    \colhead{ Property \hspace{0.15in}} &
    \colhead{\hspace{0.15in} Value \hspace{0.15in}} &
    \colhead{\hspace{0.15in} Ref} \hspace{0.35in}
}
\startdata
$P_{orb, AB}$ & 1.568024 d & a\\
$t_{0, AB}$ & 2441374.5954 JD & b\\
$P_{orb, C}$ & $190.529\pm0.061$ d & c\\
$t_{0, C}$ & $2448063.8\pm3.6$ JD & c\\
$M_A$ & $1.255\pm0.008 M_\odot$ & c\\
$M_B$ & $1.219\pm0.007 M_\odot$ & c\\
$M_C$ & $\sim$0.5 $M_\odot$ & c\\
$T_{eff,A}$ & $6500\pm50$ K & c\\
$T_{eff,B}$ & $6400\pm50$ K & c\\
$R_{A}$ & $1.2747\pm0.0072 R_\odot$ & c\\
$R_{B}$ & $1.2161\pm0.0071 R_\odot$ & c\\
$dist$ & $103\pm7$ pc & d \\
$P_{nodal}$ & $1194\pm20$ years & e\\
$i_{crit}$ & $71.3^\circ$ & e\\
$R_2 / R_1$ & 0.934 & e \\
$R_2 + R_1$ & 2.455 & e \\
$F_2 / F_1$ & 0.800 & e \\
$e \sin \omega$ & 0.00 & e \\
$e \cos \omega$ & 0.00 & e 
\enddata
\tablenotetext{a}{ \citep{popper1971}}
\tablenotetext{b}{ \citep{gyldenkerne1975}}
\tablenotetext{c}{ \citep{torres1997}}
\tablenotetext{d}{ \citep{gaia_edr3}}
\tablenotetext{e}{ This Work}
\end{deluxetable}

\section{Similar Systems with TESS}
\label{sec:others}

Aside from HS Hya, several other inclination-changing EB systems are known that can be studied with TESS. Along with surveying the Magellenic Clouds for EBs whose eclipse depths change over time, \citet{jurysek2018} provided the most comprehensive list of well characterized inclination-changing EBs within the Milky Way. A total of 11 systems, including HS Hya, have been discovered from ground-based data around bright stars. Thanks to its unmatched photometric precision and long time baseline, Kepler also discovered 42 additional inclination changing EBs in the Milky Way \citep{borkovits2016}.

To illustrate the utility in studying such systems with TESS, we searched for available TESS data for the 10 additional systems listed in Table 1 of \citet{jurysek2018}. For each system we used {\tt Eleanor} \citep{feinstein2019} to produce a 30-minute cadence light curve from the available Full Frame Image data from TESS Cyles 1 and 2 (Sectors 001--0026). Only one of the 11 systems did not have coverage from the current available TESS imaging, SV Gem. This is approximately the yield expected for TESS, now that the mission has imaged $\sim$80\% of the sky.

Many of the systems have clear eclipses still visible in the {\tt Eleanor} light curves, including: RW Per, IU Aur, AH Cep, and V685 Cen. Three of the systems had either no clear sign of eclipses in their light curves, or the data were too noisy to identify eclipses in this by-eye analysis: V669 Cyg, V907 Sco, and SS Lac. Of these, only SS Lac has been previously found to no longer eclipse. The TESS light curve for AY Mus shows $<$1\% sinusoidal variability at the previously published orbital period listed by \citet{jurysek2018}, but no clear sign of eclipse, indicating this system is also now a former EB.

A periodogram analysis for QX Cas, which was previously reported to no longer have eclipses \citep{bonaro2009}, shows very weak modulation at the published orbital period (P=6.00 days). This weak signal is likely due to the ellipsoidal modulations of QX Cas. Interestingly, the TESS light curve for QX Cas has two additional periodic signals: starspot-like sinusoidal modulation at a shorter period (1.7 day), as well as an eclipse signature with a period of 1.04 day. As the TESS pixel scale is quite large, it is entirely possible these new periodic signals may be due to contamination from a background star. However, the nearest star (TIC 2047793341) is 3.66'' away, and $\sim$24 times (3.4 mag) fainter. This makes it unclear how the relatively high amplitude spot modulations observed (1.5\% flux modulations) could be due to such a faint background star, unless the fainter star had {\it very} larger amplitude spots. It is also unclear if the P=1.04 day eclipse period is associated with this background star. If instead the new rotation and eclipse signatures are from the QX Cas system, then we speculate they may be due to the tertiary stellar component. This interesting system clearly warrants further analysis with higher spatial resolution facilities.

This brief exploration of known and bright systems highlights the potential for TESS to revolutionize the study of inclination-changing systems within the Milky Way. The excellent photometric precision of TESS means that systems with grazing eclipses such as HS Hya can be easily detected. As TESS continues into its first extended mission, the years-long baseline means that tiny changes in eclipse depth can be discovered, even for systems with potentially a 1000-year nodal period, as demonstrated by Kepler \citep{borkovits2016}. The nearly full-sky coverage of TESS means that analysis of millions of bright stars within the Full Frame Image data will almost certainly yield the largest homogeneous catalog of EBs and inclination-changing EBs ever assembled. As the stars studied by TESS are relatively bright, such EBs are also ideal for ground-based follow-up spectroscopic characterization from moderate aperture facilities.

\section{Discussion}
\label{sec:discuss}

We have examined data for the eclipsing binary HS Hydrae spanning a remarkable baseline of +125 years. This system is remarkable both in that it belongs to the rare class of precessing or inclination-changing binaries, and also in that its entire history as an eclipsing binary has been recorded in the modern era. The range of data analyzed here -- from photographic plates to space-based cameras -- shows the power in deep archival astronomy on the century timescale. We believe TESS will be a revolution for such long-duration projects, and may indeed lay the foundation for archival research of bright stars for the {\it next} century to come.

As we have noted in \S\ref{sec:others}, TESS will be an excellent platform for discovering and tracking inclination-changing EBs, with nearly all the known systems within the Milky Way already having precision data from TESS Cycles 1 and 2 available. 
As HS Hya continues to change orientation, subsequent visits by TESS should be able to track the decline of the ellipsoidal variation, however. Excitingly, TESS should be able to detect slow, monotonic increases in ellipsoidal variations, which may become {\it future} EBs.

Finally we note that after submission of this manuscript, TESS again observed the HS Hya system in Sector 35 (2021-Feb-09 to 2021-Mar-07). This is quite close to our predicted date for the system reaching the critical inclination (i.e. eclipses ceasing to be observable). As expected from our modeling of the inclination precession in \S\ref{sec:final}, no eclipses are now visible in the Sector 35 light curve. HS Hya therefore has now indeed become a {\it former} eclipsing binary.

\begin{acknowledgements}

The authors wish to thank the anonymous referee and journal editing team, whose timely and helpful suggestions substantially improved this manuscript. 

The authors wish to thank Edward Guinan and Petr Zasche for their helpful discussions about HS Hya and QX Cas. We  thank Trevor Dorn-Wallenstein, Tyler Gordon, Guadalupe Tovar Mendoza, Spencer Wallace, and Kevin R. Covey for their ongoing discussions and collaboration in analyzing TESS data. JRAD wishes to acknowledge the late Roger Griffin, whose work on long-duration binary monitoring helped inspire this work \citep{griffin2012}.

JRAD gratefully acknowledges support from the Heising-Simons Foundation, and from the Research Corporation for Science Advancement for hosting the 2019 Scialog meeting on Time Domain Astrophysics. 
JRAD also acknowledges support from the DIRAC Institute in the Department of Astronomy at the University of Washington. The DIRAC Institute is supported through generous gifts from the Charles and Lisa Simonyi Fund for Arts and Sciences, and the Washington Research Foundation. 

The work of Erin L. Howard was supported in part by the Distributed Research Experiences for Undergraduates (DREU) program, a joint project of the CRA Committee on the Status of Women in Computing Research (CRA-W) and the Coalition to Diversify Computing (CDC), which is funded in part by the NSF Broadening Participation in Computing program (NSF BPC-A \#1246649).

Courtney Klein was supported by a National Science Foundation Graduate Research Fellowship Program under grant DGE-1839285.

Jessica Birky was supported by a National Science Foundation Graduate Research Fellowship Program under grant DGE-1762114.

The DASCH project at Harvard is grateful for partial support from NSF grants AST-0407380, AST-0909073, and AST-1313370. 
This paper includes data collected by the TESS mission, which are publicly available from the Mikulski Archive for Space Telescopes (MAST). Funding for the TESS mission is provided by NASA's Science Mission directorate.
\end{acknowledgements}

\software{Python, IPython \citep{ipython}, 
NumPy \citep{numpy}, 
Matplotlib \citep{matplotlib}, 
SciPy \citep{scipy}, 
Pandas \citep{pandas}, 
Astropy \citep{astropy}, 
Lightkurve \citep{lightkurve}, 
emcee \citep{emcee}, 
eleanor \citep{feinstein2019}
}

\end{document}